# Controlling relaxation dynamics of excitonic states in monolayer transition metal dichalcogenides $WS_2$ through interface engineering

Anran Wang[1,2], Yuhan Wang[1,2], Jianfei Li[1,2], Ning Xu[1], Songlin Li[1], Xinran Wang[1], Yi Shi[1], and Fengqiu Wang[1,2*]

[1]School of Electronic Science and Engineering and Collaborative Innovation Center of Advanced Microstructures, Nanjing University, Nanjing 210093, China
[2]Key Laboratory of Intelligent Optical Sensing and Manipulation, Ministry of Education, Nanjing University, Nanjing 210093, China
Emails: fwang@nju.edu.cn

**Abstract**

Transition metal dichalcogenides (TMDs) are known to support complex excitonic states. Revealing the differences in relaxation dynamics among different excitonic species and elucidating the transition dynamics between them may provide important guidelines for designing TMD-based excitonic devices. Combining photoluminescence (PL) and reflectance contrast measurements with ultrafast pump-probe spectroscopy under cryogenic temperatures, we herein study the relaxation dynamics of neutral and charged excitons in a back-gate-controlled monolayer device. Pump-probe results reveal quite different relaxation dynamics of excitonic states under different interfacial conditions: while neutral excitons experience much longer lifetime than trions in monolayer $WS_2$, the opposite is true in the $WS_2$/h-BN heterostructure. It is found that the insertion of h-BN layer between the TMD monolayer and the substrate has a great influence on the lifetimes of different excitonic states. The h-BN flakes can not only screen the effects of impurities and defects at the interface, but also help establish a non-radiative transition from neutral excitons to trions to be the dominant relaxation pathway, under cryogenic temperature. Our findings highlight the important role interface may play in governing the transient properties of carriers in 2D semiconductors, and may also have implications for designing light-emitting and photo-detecting devices based on TMDs.

Atomically thin transition metal dichalcogenides (TMDs) have been widely studied because of their unique electrical and optical properties [1-4]. Due to reduced dielectric screening and enhanced Coulomb interactions, TMDs exhibit large binding energies, high quantum yield and controllable fluorescence efficiency [5-9]. Besides excitons, TMDs and Van der Waals heterostructures have been confirmed to support abundant multi-particle excitonic states including trions, biexcitons, exciton-trion complexes and interlayer excitons, providing an ideal platform to study two-dimensional excitonic physics [9-12]. However, electrical and optical properties of TMDs can be susceptible to extrinsic effects [13-16]. Simple yet effective approaches for customizing the lifetimes of various excitonic states in TMDs is still lacking. Recently, different oxide substrates are found to strongly affect the relaxation lifetimes of excitons in monolayer $MoSe_2$ due to the interlayer electron-phonon coupling [14]. Moreover, the competition between radiative and non-radiative relaxation pathways of excitons has been manipulated by background carrier concentration via electrostatic doping in a field-effect device, pinpointing the importance of the interplay between excitons, trions and free-carriers [17]. Therefore, a comprehensive and in-depth understanding of the transition dynamics between these quasiparticles is becoming an important topic.

While there have been a number of theoretical works addressing the carrier concentration and temperature dependence of radiative lifetimes for excitons and trions [18, 19], influence of the interaction between excitons, trions and free-carriers has not been taken fully into account. Recently, the interconversion between excitons and trions was investigated as the background carrier concentration was tuned. Suppression of non-radiative relaxation pathways resulted in a near-unity quantum yield for monolayer $MoS_2$ at room temperature [17]. Time-resolved PL spectroscopy finds that the exciton lifetime decreases with increasing electron concentration for monolayer $MoS_2$ at room temperature, indicating the increased probability of negative trion formation [20]. While the above works confirm the interaction among different carrier species, the transition dynamics of such interaction has not been investigated thoroughly. On one hand, the energies of neutral excitons and trions cannot be well resolved at room temperature, making it experimentally difficult to study the intrinsic features of individual transitions; on the other hand, the transitions between different species are typically of non-radiative nature, which can be more suitably probed by transient absorption spectroscopy.

In this study, we fabricate back-gated field-effect transistors based on WS$_2$ monolayer with different interfacial conditions, where the static optical properties and the transient dynamics of neutral and charged excitons can be conveniently measured under different back-gate voltages. The experiments were carried out at low temperatures so that energies of excitons and trions can be well resolved, and the interaction between excitons and phonons can be blocked. We use a h-BN passivation layer to isolate the sample from the substrate, and the influence of doping and localization from interfacial defect states can be controlled. Our results show very distinguishable transient dynamics for excitons and trions, where the neutral excitons follow a very fast decay in the few picosecond range, but the lifetime of trions is in the hundreds of picosecond range in WS$_2$/h-BN heterostructure. In particular, it is found that the insertion of h-BN flakes greatly affects the dynamics of neutral and charged excitons, resulting in the opposite lifetime features observed in the monolayer WS$_2$ in direct contact with the substrate. The ultrafast exciton-to-trion transition serves as the dominant non-radiative relaxation pathway for neutral excitons in WS$_2$/h-BN samples at low temperature. Our results provide clear evidence for the dynamical coupling between neutral excitons, trions and free-carriers, and demonstrate the importance of interfacial engineering in tailoring photocarrier lifetimes and designing TMD-based devices.

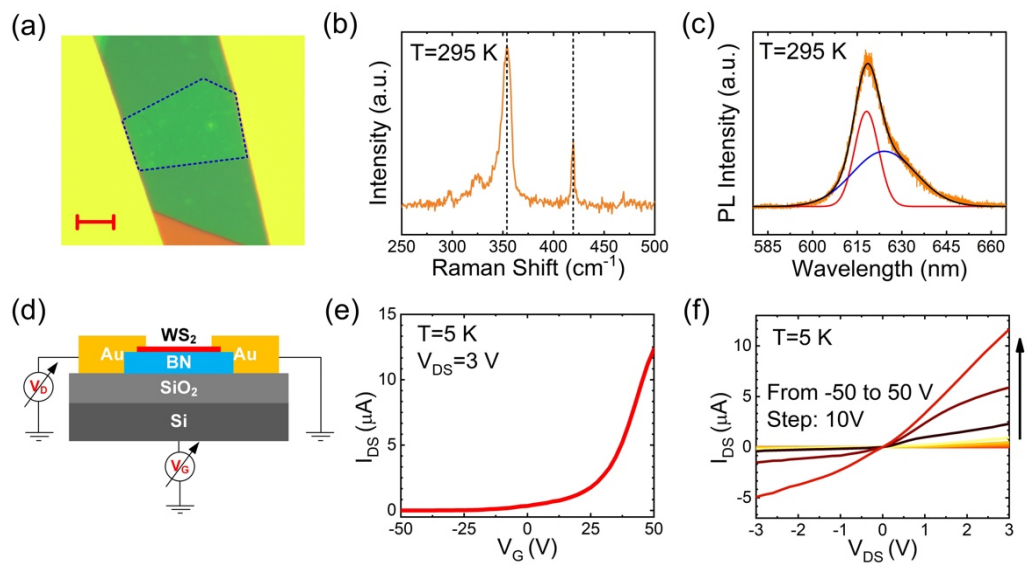

Figure 1(a) Optical microscope image of the WS$_2$/h-BN heterostructure. The scale bar is 10 μm. (b) Raman and (c) photoluminescence spectra of the monolayer WS$_2$ at room temperature. The orange curve corresponds to the experimental data. The black curve in (c) corresponds to the total emission of monolayer, and the red and blue curves correspond to the emission of neutral excitons and trions by Gaussian fitting, respectively. (d) Schematic illustration of the back-gate-controlled WS$_2$/h-BN device. (e) Transfer curve of the device at 5 K. (f) Output curves of the device under different gate

voltages from -50 V to 50 V.

WS$_2$ monolayer and h-BN flakes were fabricated by mechanical exfoliation from the respective single crystals, and transferred onto SiO$_2$(285 nm)/Si substrates to form the WS$_2$/h-BN heterostructure, as shown by the optical microscope image in Fig. 1a. The thickness of the h-BN flake underneath the WS$_2$ layer is ~32 nm as determined by atomic force microscope. The relatively thick h-BN flake used here serves as a passivation layer to mitigate the influence of the charged impurities and defects from the oxide substrates and the interface, so that the intrinsic fine structures of monolayer WS$_2$ can be better evaluated. As shown in Fig. 1b, two dominant peaks at 354.2 cm$^{-1}$ and 419.4 cm$^{-1}$ in Raman spectrum are observed, which correspond to the in-plane $E_{2g}^1$ and out-of-plane A$_{1g}$ optical modes of WS$_2$ [21]. The monolayer nature of the marked area on the WS$_2$ flake was confirmed by PL spectroscopy at room temperature with 514 nm (2.41 eV) excitation. In Fig. 1c, the PL spectrum with a sharp peak located at around 618 nm (~2.00 eV) is consistent with the direct bandgap of monolayer WS$_2$ at room temperature [21, 22]. The asymmetric shape of the peak can be attributed to the emission of trions due to unintentional *n*-doping during the growth and fabrication process of samples, and the peak position of trions is around 624 nm (~1.98 eV), as estimated by Gaussian fitting.

To investigate the influence of background carrier doping concentration, we fabricate a back-gate-controlled field-effect transistor, as illustrated in Fig. 1d. The transfer and output curves are shown in Fig. 1e and 1f. The output curves under different back-gate voltages confirm effective tuning by back-gating on the background carrier concentration in WS$_2$. The induced carrier concentration $n_e$ by the back-gate can be calculated by the relation $n_e = C_g(V_G - V_{th})/e$, where $V_G$ is the bake-gate voltage, $V_{th}$ is the threshold voltage, $e$ is the electronic charge and $C_g$ is the combined gate capacitance [23]. For the h-BN flakes with a thickness of 32 nm and the SiO$_2$ film of 285 nm, the combined gate capacitance is ~1.09×10$^{-8}$ F/cm$^2$, and the induced doping carrier concentration can be evaluated to be ~1.64×10$^{12}$ cm$^{-2}$ at 50 V. Also, we have fabricated controlled back-gated devices of monolayer WS$_2$ deposited on SiO$_2$/Si substrate to compare the effects of different interfaces (See Supplementary).

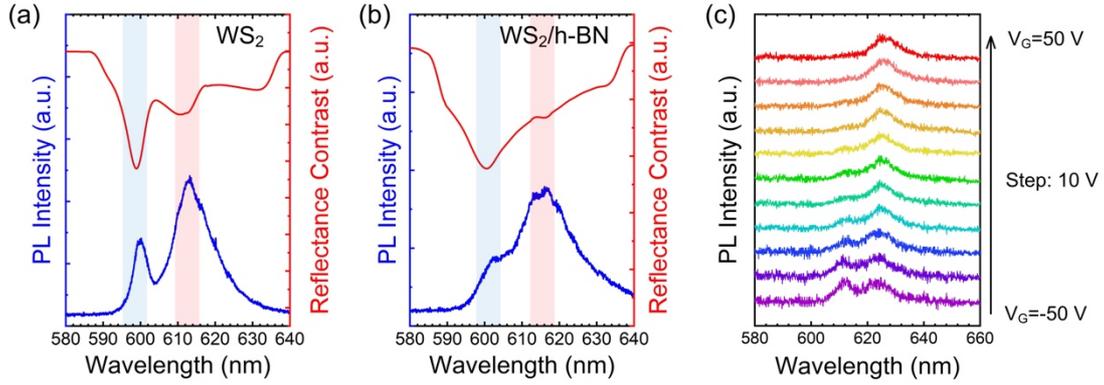

Fig 2 PL and reflectance contrast spectra of (a) monolayer WS$_2$ and (b) WS$_2$/h-BN at 150 K. The red and blue stripes are visual guides marking the peak positions of the reflectance contrast and PL spectrum, respectively. (c) PL spectra of annealed WS$_2$/h-BN heterostructure under gate voltages from -50 V to 50 V at 5 K.

To identify the excitonic features, we performed PL and reflectance contrast spectra of monolayer WS$_2$ and WS$_2$/h-BN heterostructure at a relative low temperature, where both excitons and trions have appreciable oscillator strength. Multiple samples are prepared and characterized, and typical results are shown in Fig. 2, where a clear correlation between the absorption and the emission peaks of excitons can be seen. We used a supercontinuum light source to obtain the reflectance contrast spectrum. The average power is carefully controlled not to induce saturable absorption. Typically, the neutral excitons exhibit stronger absorption features than trions. Compared with a monolayer WS$_2$, the emission and reflectance peaks of WS$_2$ deposited on the h-BN flake are red-shifted, revealing the obvious influence, *i.e.,* reduced binding energy, of the dielectric environment on the excitonic states [24, 25]. The clearly broadened emission and absorption linewidths for excitons, as shown in Fig.2, suggest there are mechanisms affecting the dephasing time or the exciton relaxation channels [26].

To get better interlayer coupling, the WS$_2$/h-BN heterostructure is annealed to optimize the contact of the two flakes. PL spectra of the WS$_2$/h-BN heterostructure under various gate voltages is shown in Fig. 2c. In contrast to the PL spectrum at room temperature, the emission of neutral excitons (~2.02 eV) and trions (~1.98 eV) can be well resolved, and the binding energy for trions ~40 meV is consistent with previously reported values in Ref. [27-29]. To investigate the effect of background carriers, the PL spectra were measured as the back-gate voltage varied from -50 V to 50 V at 5 K. As the back-gate voltage increases, the trion peak position is slightly red-shifted, which is as a result of the reduced trion binding energy due to enhanced carrier screening effect from induced carriers by the back-gate voltage [30]. Moreover, the intensity of neutral exciton emission weakens rapidly with increase voltage, but the trion emission increases, agreeing with the assignment of the two PL peaks to excitons and trions. The presence of the trion emission under all applied back-gate voltages indicates heavy n-doping of the $WS_2$ layer. The well-resolved PL and absorption peaks make it experimentally feasible to trace distinct excitonic states and study their transient dynamics.

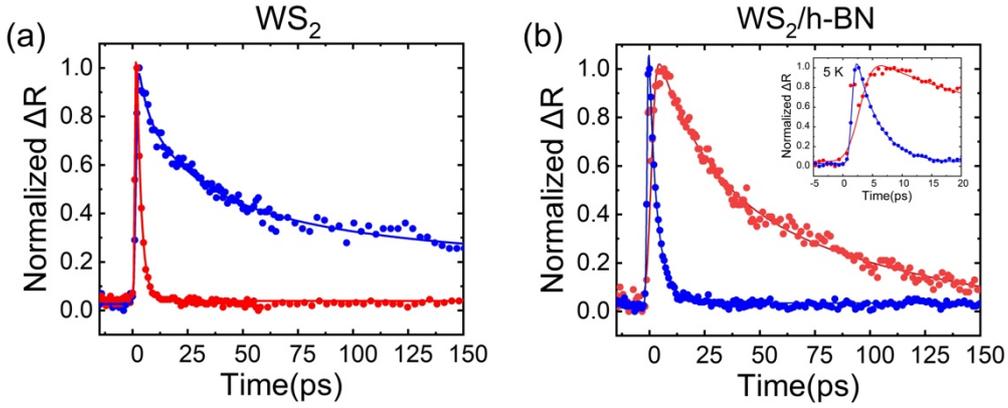

Fig 3 Normalized transient reflectance of neutral excitons (blue) and trions (red) measured at zero back-gate voltage of annealed (a) monolayer $WS_2$ and (b) $WS_2$/h-BN heterostructure at 5 K. Inset: the normalized transient reflectance in a short time scale. All the measurements were performed at back-gate voltage of 0 V.

To resolve the relaxation dynamics of neutral and charged excitons, we performed ultrafast pump-probe under zero back-gate voltage. For the purpose of ruling out the influence of phonons, all the measurements were carried out at 5K. The monolayer $WS_2$ was excited with a pump wavelength of 400 nm (3.10 eV), and probed at the resonant emission peaks of neutral excitons and trions, respectively. It should be noted that the linewidth of the femtosecond probe beam (~29 meV) is appreciably smaller than the trion binding energy (~40 meV), which enables unambiguous tracking of photocarriers dynamics at closely spaced energy states (See Supplementary). In our experiments, we mainly focus on the transient signals probed at resonance wavelengths with excitons and trions. The applied pumping fluence is ~18 μJ/cm$^2$, and the corresponding injected carrier concentration is ~1.7×10$^{12}$ cm$^{-2}$. The normalized transient reflectance of these two kinds of excitonic states in annealed sample is shown in Fig. 3.

In contrast to the case of monolayer $WS_2$ directly deposited on $SiO_2$/Si substrate, the inserted h-BN layer has led to significant modification on the transient behaviors of these two excitonic species (See Supplementary). For monolayer WS$_2$ directly deposited on the SiO$_2$/Si substrate, trions are found to decay immediately after the pump as shown in Fig. 3a, due to fast trapping by charged impurities or defects at the WS$_2$/SiO$_2$ interface, resulting in a quick non-radiative relaxation. Whereas the electrically neutral nature of excitons makes it less susceptible to impurities or traps, and thus neutral excitons exhibit a slower decay process compared to trions [17].

But for the passivated WS$_2$/h-BN heterostructure, the recombination lifetimes of neutral and charged excitons are distinct from monolayer WS$_2$, where neutral excitons follow a very fast mono-exponential decay with a lifetime of 3.8±0.1 ps and trions follow a much slower bi-exponential decay at 5 K. For trions, the fast decay component $\tau$=22.8±2.2 ps can be attributed to the non-radiative relaxation process, such as the defect-assisted trapping and/or the geminate Auger-like recombination [17], where the additional charge can serve as the third particle required for momentum conservation. The longer component with $\tau$=125.6±28.9 ps corresponds to the radiative recombination lifetime relaxation process of trions. Compared with neutral excitons, trions are more likely to be localized by disorder traps or defects at low-temperature due to the additional electrons and larger mass [19, 22]. According to the Pauli blocking effect, it is difficult to find unoccupied states for excess electrons from trions after recombination. As a result, trions are expected to follow a much longer radiative relaxation than neutral excitons do.

It is noticed that the rise time of trions transient reflectance signals is ~2.4±0.3 ps, in agreement with the exponential decay lifetime of neutral excitons, which suggests that the rising process of the trions transient signals is related to the transition from neutral excitons to trions (inset of Fig. 3b). Note that the reported radiative lifetime of neutral excitons in monolayer TMD is around 2-3 ps [18, 29], close to the decay lifetime of 3.8±0.1 ps in our samples. However, considering the low PL intensity of neutral excitons even at low temperature in our samples, the fast decay of neutral excitons is unlikely to be related to the radiative recombination.

In addition, to study the transition between neutral excitons and trions, we have measured the transient dynamics of two excitonic states at different temperatures. The transitions can be observed up to the temperature of liquid nitrogen ~80 K (See Supplementary). With increasing temperature, the weight of phonon-assisted relaxation processes increases, inducing an additional non-radiative relaxation pathway of neutral excitons. Moreover, the thermal instability of trions makes it difficult to reveal the relaxation dynamics of trions at higher temperatures [13].

It should be noted that the contrasting dynamics between monolayer $WS_2$ and $WS_2$/h-BN heterostructure arise simply from the use of a h-BN passivation layer. On one hand, the thick h-BN flake help screen the influence of disorder potential fluctuations and Coulomb traps [31, 32]. Within the allowed pump fluence range (varying from 18 to 36 μJ/cm$^2$), we have not observed any significant change of neutral exciton lifetimes, suggesting the limited influence of defects- and/or traps-assisted relaxation pathways. On the other hand, the enhanced dielectric screening and reduced binding energy leads to enhanced exciton diffusion [33] and increased exciton Bohr radius [34], which help facilitate the interaction between different excitonic species. Therefore, the non-radiative process such as the transition from neutral excitons to trions is likely to be the primary relaxation pathway of neutral excitons in $WS_2$/h-BN heterostructure.

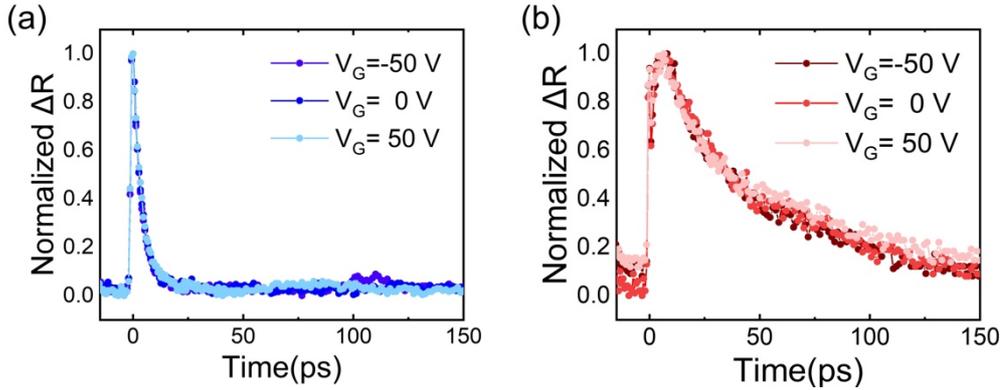

Fig 4 Normalized transient reflectance of neutral excitons (a) and trions (b) measured under different back-gate voltages. All measurements are performed at 5 K.

To further investigate the relaxation and transition mechanism, the pump-probe spectra of neutral and charged excitons under different back-gate voltages at 5 K was measured in Fig. 4a and 4b. With the back-gate voltage varying from -50 V to 50 V, the transition time from neutral excitons to trions and the lifetimes for both two excitonic species do not show appreciable changes. The limited modulation of background carrier concentration induced by electrostatic doping may be the main reason. Employing chemical doping or the combination of both are expected to induce much higher doping concentration and are more likely to show observable impacts on dynamics of photocarriers [20, 35].

To summarize, by photoluminescence and reflectance spectroscopy, we have clearly resolved the very distinguished lifetimes of neutral and charged excitons at cryogenic temperature. The use of h-BN passivation layer is seen to significantly alter the ultrafast response from different excitonic species. An ultrafast transition from neutral exciton to trion becomes the dominant fast non-radiative relaxation pathway at low-temperature in strongly coupled $WS_2$/h-BN heterostructure, which is very different from the exciton dynamics in monolayer $WS_2$ deposited directly on $SiO_2$/Si substrate. Electrostatic doping induced by the back gate (~ $1.64 \times 10^{12}$ $cm^{-2}$ at 50 V) is found to have negligible effects on the relaxation dynamics of the two excitonic species. Our findings have implications for fabricating light-emitting devices and photo-detectors based on TMD materials.

**Supplementary Material**

See supplementary material for electrical and optical characterizations of monolayer $WS_2$ sample, as well as temperature-dependent pump-probe results obtained for the $WS_2$/h-BN heterostructure.

**Acknowledgements**

This work was supported in part by the State Key Project of Research and Development of China (2017YFA0206304, 2018YFB2200500); National Natural Science Foundation of China (61775093, 61427812); National Youth 1000-Talent Plan; A 'Jiangsu Shuangchuang Team' Program; Natural Science Foundation of Jiangsu Province (BK20170012).

**Data availability**

The data that support the findings of this study are available from the corresponding author upon reasonable request.

# References

[1] Mak, K.F., Lee, C., Hone, J., Shan, J., Heinz, T.F. Atomically thin $MoS_2$: a new direct-gap semiconductor. Phys. Rev. Lett. 105, 136805 (2010).

[2] Radisavljevic, B., Radenovic, A., Brivio, J., Giacometti, V., Kis, A. Single-layer $MoS_2$ transistors. Nat. Nanotechnol. 6, 147–150 (2011).

[3] Wang, Q., Kalantar-Zadeh, K., Kis, A., Coleman, J.N., Strano, M.S. Electronics and optoelectronics of two-dimensional transition metal dichalcogenides. Nat. Nanotechnol. 7, 699–712 (2012).

[4] Nie, Z., Wang, Y., Li, Z., Sun, Y., Qin, S., Liu, X., Turcu, I.C.E., Shi, Y., Zhang, R., Ye, Y. et al. Ultrafast free carrier dynamics in black phosphorus-molybdenum disulfide (BP/$MoS_2$) heterostructure. Nanoscale Horizons. (2019).

[5] Chernikov, A., Berkelbach, T.C., Hill, H.M., Rigosi, A., Li, Y., Aslan O.B., Reichman, D.R., Hybertsen, M.S., Heinz, T.F. Exciton binding energy and nonhydrogenic Rydberg series in monolayer $WS_2$. Phys. Rev. Lett. 113, 076802 (2014).

[6] Ugeda, M.M., Bradley, A.J., Shi, S., da Jornada, F.H., Zhang, Y., Qiu, D., Ruan, W., Mo, S., Hussain, Z., Shen, Z. et al. Giant bandgap renormalization and excitonic effects in a monolayer transition metal dichalcogenide semiconductor. Nat. Mater. 13, 1091–1095 (2014).

[7] Amani, M., Lien, D., Kiriya, D., Xiao, J., Azcatl, A., Noh, J., Madhvapathy, S.R., Addou, R., KC, S., Dubey, M. et al. Near-unity photoluminescence quantum yield in $MoS_2$. Science 350, 1065–1068 (2015).

[8] Li, Y., Liu, W., Xu, H., Chen, H., Ren, H., Shi, J., Du, W., Zhang, W., Feng, Q., Yan, J. et al. Unveiling bandgap evolution and carrier redistribution in multilayer $WSe_2$: enhanced photon emission via heat engineering. Advanced Optical Materials, 8, 1901226 (2020)

[9] Mak, K.F., He, K., Lee, C., Lee, G.H., Hone, J., Heinz, T.F., Shan, J. Tightly bound trions in monolayer $MoS_2$. Nat. Mater. 12, 207–211 (2013).

[10] Barbone, M., Montblanch, A.R.P., Kara, D.M., Palacios-Berraquero, C., Cadore, A.R., De Fazio, D., Pingault, B., Mostaani, E., Li, H., Chen, B. et al. Charge-tuneable biexciton complexes in monolayer $WSe_2$. Nat. Commun. 9, 3721 (2018).

[11] Chen, S.Y., Goldstein, T., Taniguchi, T., Watanabe, K., Yan, J. Coulomb-bound four- and five-particle intervalley states in an atomically-thin semiconductor. Nat. Commun. 9, 3717 (2018).

[12] Shi, J., Li, Y., Zhang, Z., Feng, W., Wang, Q., Ren, S., Zhang, J., Du, W., Wu, X., Sui, X. et al. Twisted-angle-dependent optical behaviors of intralayer excitons and trions in $WS_2$/$WSe_2$ heterostructure. ACS Photonics 6, 3082 (2019).

[13] Ross, J.S., Wu, S., Yu, H., Ghimire, N.J., Jones, A.M., Aivazian, G., Yan, J., Mandrus, D. G., Xiao, D., Yao, W. et al. Electrical control of neutral and charged excitons in a monolayer semiconductor. Nat. Commun. 4, 1474 (2013).

[14] Nie, Z., Shi, Y., Qin, S., Wang, Y., Jiang, H., Zheng, Q., Cui, Y., Meng, Y., Song, F., Wang, X. et al. Tailoring exciton dynamics of monolayer transition metal dichalcogenides by interfacial electron-phonon coupling. Commun. Phys. 2, 103 (2019).

[15] Wang, H., Zhang, C., Rana, F. Ultrafast dynamics of defect-assisted electron–hole recombination in monolayer $MoS_2$. Nano Letters 15, 339-345 (2015).

[16] Shi, J., Zhu, J., Wu, X., Zheng, B., Chen, J., Sui, X., Zhang, S., Shi, J., Du, W., Zhong, Y. et al. Enhanced trionemission and carrier dynamics in monolayer $WS_2$ coupled with plasmonic nanocavity. Advanced Optical Materials, 8, 2001147 (2020)

[17] Lien, D., Uddin, S.Z., Yeh, M., Amani, M., Kim, H., Ager III, J.W., Yablonovitch, E., Javey, A. Electrical suppression of all nonradiative recombination pathways in monolayer semiconductors. Science 364, 468 (2019).

[18] Wang, H., Zhang, C., Chan, W., Manolatou, C., Tiwari, S., Rana, F. Radiative lifetimes of excitons and trions in monolayers of the metal dichalcogenide $MoS_2$. Phys. Rev. B 93, 045407 (2016).

[19] Steinhoff, A., Rösner, M., Jahnke, F., Wehling, T.O., Gies, C. Influence of excited carriers on the optical and electronic properties of $MoS_2$. Nano letters 14, 3743–3748 (2014).

[20] Pradeepa, H.L., Praloy, M., Aveek, B., Jaydeep, K.B. Electrical and chemical tuning of exciton lifetime in monolayer $MoS_2$ for Field-Effect Transistors. ACS Appl. Nano Mater. 3, 641–647 (2020).

[21] Shi, W., Lin, M., Tan, Q., Qiao, X., Zhang, J., Tan, P. Raman and photoluminescence spectra

of two-dimensional nanocrystallites of monolayer $WS_2$ and $WSe_2$. 2D Materials, 3, 025016 (2016).

[22] Gutierrez H.R., Perea-López, N., Laura Elías, A., Berkdemir, A., Wang, B., Lv, R., López-Urías, F., Crespi, V.H., Terrones, H., Terrones, M. Extraordinary Room-Temperature Photoluminescence in Triangular $WS_2$ Monolayers. Nano Letters, 13, 3447 (2013).

[23] Aftab, S., Akhtar, I., Seo, Y., Eom, J. $WSe_2$ homojunction p-n diode formed by photoinduced activation of mid-gap defect states in boron nitride. ACS Appl. Mater. Interfaces (2020).

[24] Raja, A., Chaves, A., Yu, J., Arefe, G., Hill, H.M., Rigosi, A.F., Berkelbach, T.C., Nagler, P., Schüller, C., Korn, T. et al. Coulomb engineering of the bandgap and excitons in two-dimensional materials. Nat Commun 8, 15251 (2017).

[25] Pogna, E.A.A., Marsili, M., De Fazio, D., Dal Conte, S., Manzoni, C., Sangalli, D., Yoon, D., Lombardo, A., Ferrari, A.C., Marini, A. et al. Photo-Induced Bandgap Renormalization Governs the Ultrafast Response of Single-Layer $MoS_2$. ACS Nano, 10, 1182 (2016).

[26] Mueller, T., Malic, E. Exciton physics and device application of two-dimensional transition metal dichalcogenide semiconductors. npj 2D Mater Appl 2, 29 (2018).

[27] Wang, G., Bouet, L., Lagarde, D., Vidal, M., Balocchi, A., Amand, T., Marie, X., Urbaszek, B. Valley dynamics probed through charged and neutral exciton emission in monolayer $WSe_2$. Phys. Rev. B 90, 075413 (2014).

[28] Fang, H.H., Han, B., Robert, C., Semina, M.A., Lagarde, D., Courtade, E., Taniguchi, T., Watanabe, K., Amand, T., Urbaszek, B. et al. Control of the exciton radiative lifetime in van der Waals heterostructures. Phys. Rev. Lett. 123, 067401 (2019).

[29] Robert, C., Lagarde, D., Cadiz, F., Wang, G., Lassagne, B., Amand, T., Balocchi, A., Renucci, P., Tongay, S., Urbaszek, B. et al. Exciton radiative lifetime in transition metal dichalcogenide monolayers. Phys. Rev. B 93, 205423 (2016).

[30] Chernikov, A., van der Zande, A.M., Hill, H.M., Rigosi, A.F., Velauthapillai, A., Hone, J., Heinz, T.F. Electrical Tuning of Exciton Binding Energies in Monolayer $WS_2$. Phys. Rev. Lett. 115, 126802 (2015).

[31] Katoch, J., Ulstrup, S., Koch, R.J., Moser, S., McCreary, K.M., Singh, S., Xu, J., Jonker, B.T., Kawakami, R.K., Bostwick, A. et al. Giant spin-splitting and gap renormalization driven by trions in single-layer $WS_2$/h-BN heterostructures. Nature Phys. 14, 355–359 (2018).

[32] Hoshi, Y., Kuroda, T., Okada, M., Moriya, R., Masubuchi, S., Watanabe, K., Taniguchi, T., Kitaura, R., Machida, T. Suppression of exciton-exciton annihilation in tungsten disulfide monolayers encapsulated by hexagonal boron nitrides. Phy. Rev. B, 95, 241403 (2017).

[33] Kang, J., Jung, J., Lee, T., Kim, J.G., Cho, C. Enhancing exciton diffusion in monolayer $WS_2$ with h-BN bottom layer. Phys. Rev. B 100, 205304 (2019).

[34] Bastard, G., Mendez, E.E., Chang, L.L., Esaki, L. Exciton binding energy in quantum wells. Phys. Rev. B 26, 1974 (1982).

[35] Sun, Y., Meng, Y., Dai, R., Yang, Y., Xu, Y., Zhu, S., Shi, Y., Xiu, F., Wang, F. Slowing down photocarrier relaxation in Dirac semimetal $Cd_3As_2$ via Mn doping. Optics Letters, 44, 4103-4106 (2019).